\documentclass[11pt,preprint]{aastex}
\usepackage{array}
\newcolumntype{?}{!{\vrule width 1pt}}

\begin{document}

\title{A re-assessment of the Kuiper belt size distribution for sub-kilometer objects, revealing collisional equilibrium at small sizes} 

\author{\textbf{A. Morbidelli$^{(1)}$,  D. Nesvorny$^{(2)}$, W.F. Bottke$^{(2)}$, S. Marchi$^{(2)}$}\\  
(1) Laboratoire Lagrange, UMR7293, Universit\'e de Nice Sophia-Antipolis,
  CNRS, Observatoire de la C\^ote d'Azur. Boulevard de l'Observatoire,
  06304 Nice Cedex 4, France. (Email: morby@oca.eu / Fax:
  +33-4-92003118) \\
  (2) Souwthwest Research Institute, Boulder, Co.\\
} 

\accepted{Dec 3, 2020 in Icarus}
\begin{abstract}
  {\footnotesize\singlespace In this work we combine several constraints provided by the crater records on Arrokoth and the worlds of the Pluto system to compute the size-frequency distribution (SFD) of the crater production function for craters with diameter $D\lesssim 10$~km. For this purpose, we use a Kuiper belt objects (KBO) population model calibrated on telescopic surveys, that describes also the evolution of the KBO population during the early Solar System. We further calibrate this model using the crater record on Pluto, Charon and Nix.  Using this model, we compute the impact probability on Arrokoth, integrated over the age of the Solar System. {This probability is then used together with other observational constraints to determine the slope of the crater-production function on Arrokoth. These constraints are: (i) the spatial density of sub-km craters, (ii) the absence of craters with $1<D<7$~km; (iii)  the existence of a single crater with $D>7$~km.   In addition, we use our Kuiper belt model also to compare the impact rates and velocities of KBOs on Arrokoth with those on Charon, integrated over the crater retention ages of their respective surfaces. This allows us to establish a}  relationship between the spatial density of sub-km craters on Arrokoth and of $D\sim 20$~km craters on Charon. Together, all these considerations suggest the crater production function on these worlds has a cumulative power law slope of $-1.5<q<-1.2$. Converted into a projectile SFD slope, we find $-1.2<q_{\rm KBO}<-1.0$. These values are close to the cumulative slope of main belt asteroids in the 0.2--2~km range, a population in collisional equilibrium (Bottke et al. 2020). {For KBOs, however, this slope appears to extend from $\sim 2$~km down to objects a few tens of meters in diameter}, as inferred from sub-km craters on Arrokoth. From the measurement of the dust density in the Kuiper belt made by the New Horizons mission, we predict that the SFD of the KBOs becomes steep again below $\sim 10$--$30$~m. All these considerations strongly indicate that the size distribution of the KBO population is in collisional equilibrium.} \end{abstract}
%\linenumbers
\singlespace
\section{Introduction}

The {cumulative} size-frequency distribution (SFD) of Kuiper belt objects (KBO), usually expressed as a power-law of type $N(>d)\propto d^{q_{\rm KBO}}$, can only be determined { from ground-based observations for large sizes}. After some debate among competing groups, a consensus has been reached (Bernstein et al., 2004; Fuentes et al., 2009; Fraser et al., 2014; Adams et al., 2014; Emery et al., 2015) that: 
\begin {itemize}
\item The SFD is steep for objects with diameter $d>100$~km, where it has different slopes of $q_{\rm KBO}\sim -4.5$ and $-7.5$ for the hot and cold sub-populations respectively (Fraser et al., 2014). 
\item For $d\lesssim 100$~km, there is a roll-over in the SFD that becomes approximately the same for both the cold and hot sub-populations and is similar to that observed in the main asteroid belt and Jupiter's Trojans ($q_{\rm KBO}\sim-2$). 
\end {itemize} 
The directly observed SFD, however, only extends down to a few tens of km. Stellar occultations provide some insights into the nature of the KBO SFD at smaller sizes (Schlichting et al., 2012; Arimatsu et al., 2019), but limited statistics, observational biases, and uncertainties with this technique make it difficult to draw robust conclusions.

This situation changed radically with the New Horizons mission to the Pluto-Charon system and to the cold classical KBO called Arrokoth.  Superb images of their surfaces have allowed researchers to perform crater counts on KBOs for the first time. Crater-counting is a well-established technique in planetary and asteroid belt studies that can unveil the size distribution of the projectile population (see e.g. Ivanov et al., 2002 for a review). 

Crater counts on the surfaces of Pluto and Charon revealed that the KBO SFD shares reasonable similarities to that of the asteroid belt down to crater diameter $D \gtrsim 10$--$20$~km (Robbins et al., 2017, hereafter denoted R17), which corresponds to a KBO projectile size of roughly $d\gtrsim 2$~km. For sizes smaller than this value of $D$, the crater SFD appears to be very shallow (Singer et al. 2019; S19 hereafter), with an estimated cumulative slope of $q=-0.7\pm 0.3$. If confirmed, this result would imply that the KBO SFD is significantly shallower than that of the asteroid belt in the same size-range ($q_{\rm MB}=-1.29\pm 0.02$; Yoshida and Nakamura, 2007).

A possible implication of this result is that the KBO population did not reach collisional equilibrium. If true, the observed KBO SFD would presumably still be (close to) the one set by the KBO formation process and therefore could be used to test planetesimal accretion models (Johansen et al., 2015). Moreover, it would imply that the multi-km comets we observe today, such as 67P/Churyumov-Gerasimenko, are remnants of the original planetesimal population rather than fragments of larger objects (Davidsson et al., 2016). 

With that said, it is possible that the cumulative slope of $q=-0.7\pm 0.3$ is more ambiguous than suggested by the error bars. {First, Spencer et al. (2020; Sp20 hereafter) revised the slope determination as $q=-0.8^{+0.4}_{-0.6}$. Although the nominal slope is almost the same as in S19, the statistical uncertainty is much larger than originally estimated and does not exclude a slope as steep as $q=-1.4$. Second,} the slope of the crater SFD at the smallest observable sizes may also be subject to systematic uncertainties. For instance, R17 (see their Fig. 11) report $q=-1.47\pm 0.15$ for the same terrain and crater size-range considered in S19, when using craters identified in what they call the ``consensus database''. The reason for the difference between the slopes in S19 and R17 has yet to be fully investigated. One factor may be that R17 counted certain craters twice in poorly-imaged regions where there was image overlap (reviewer, personal communication).  A second factor may that S19 took a more conservative approach in accepting candidate craters near the resolution limit than R17, perhaps resulting in a shallower slope (S. Robbins, personal communication). We note that modest variability in crater counts between experts is a documented phenomenon (Robbins et al. 2014).

It is also possible that the observed crater SFD may not share the exact same shape as the projectile SFD. Not only do crater scaling laws vary widely depending on the nature of the target surface (e.g., Holsapple and Housen, 2007), but some small bodies appear to have experienced crater erasure processes (e.g., Steins, Toutatis, Itokowa, Ryugu, Bennu; see Richardson et al. 2005; Michel et al., 2009; Marchi et al. 2015; Bottke et al. 2020). {However, S19 claimed that this process would not affect their determination of the crater SFD slope.}

Fortunately, the flyby of Arrokoth by the New Horizons spacecraft has allowed researchers to obtain an independent crater count on a different KBO at crater sizes that are smaller than those observed on Pluto and Charon.  Sp20 reported that the crater SFD on Arrokoth for $0.2<D<1$~km shows a cumulative slope of $q=-1.3\pm 0.6$. This slope is steeper than that reported for Pluto and Charon in the 1--10~km $D$ range { (nominal $q=-0.7$ in S19 and $-0.8$ in Sp20)}, but it is nevertheless consistent within { the large} statistical uncertainties.  Still,  Sp20 noted that the {\it crater density on Arrokoth is higher than would be obtained from an extrapolation of the Charon slope and density to sub-km craters} (see Fig. 6B in that paper). It is possible the higher than expected crater spatial densities come from a { crater-production SFD that is steeper than that estimated from Charon's record in the range $1<D<10$~km}, but it may also have a number of other causes (e.g., different impact rates per unit surface on Arrokoth and Pluto/Charon; different relationships between projectile and crater sizes that are due to the vastly different gravities of the targets; potentially different surface ages). 

The Arrokoth crater dataset in Sp20 also presents another intriguing property. There are several craters observed that are slightly smaller than 1~km, none are found between $1<D<7$~km, and then there is one large singular $D>7$~km crater seen that has been named {\it Maryland}.  The absence of craters between the km-sized ones and Maryland is not due to limitations in the images and could suggest a relatively steep crater SFD.  We will return to this issue below.

In this paper we re-assess the underlying crater production SFD for $D\lesssim 10$~km in light of these new observations. We start in Section~\ref{ImpRates} by estimating the accumulation of impacts of projectiles with diameter $d>2$~km on the Pluto system and on Arrokoth, not only within the current Kuiper belt but also during the early evolution of the Solar System, when the current Kuiper belt took shape. In Section~\ref{MonteCarlo} we perform Monte Carlo simulations with various crater SFD slopes to assess which ones are most likely to reconcile the observed spatial density of sub-km craters on Arrokoth, the existence of the Maryland crater, the absence of intermediate-size craters and the probability of $d>2$~km impacts. In Section~\ref{offset} we convert the spatial density of sub-km craters on Arrokoth into a spatial density of equivalent craters on Charon by accounting for the ratio of impact rates on the two bodies and the different projectile-to-crater size relationships (and related uncertainties).  We use this result, together with the observed spatial density of craters with $D\sim 20$~km, to determine a crater production SFD slope on Charon that can be compared with those derived in Sect.~\ref{MonteCarlo}. Finally, in Section~\ref{conclusions}, we derive the projectile SFD from the crater SFD and discuss whether the KBO population has reached collisional equilibrium.

\section{Impact rates on Pluto/Charon and Arrokoth}
\label{ImpRates}
\subsection{Current impact rates}
\label{current}
To evaluate the { current impact rate of KBOs on the Pluto system and Arrokoth one needs to have an orbital/SFD Kuiper belt model. We use the model developed over a series of papers (Nesvorny, 2015a,b; Nesvorny and Vorkouhlicky, 2016; Nesvorny et al., 2017, 2019b). It is obtained as the final state of simulations} that describe how the primordial Kuiper belt was dynamically { sculpted} during the post-nebula migration of the giant planets.  The results were calibrated on observations from the OSSOS survey (Lawler et al., 2018), { i.e. by selecting the simulations that best reproduce in the end the currently observed belt, once observational biases are taken into account, and by tuning the number of $d>100$~km objects. For the objects between $2<d<100$~km we assume a cumulative power law distribution with a slope of $-2.1$, the same as that of the Jupiter Trojans (Grav et al., 2011; Emery et al., 2015), which are believed to be captured Kuiper belt objects (Nesvorny et al., 2013).}  In the discussion below, ``cold'' and ``hot''  are a measure of dynamical excitation, not temperature.   

Compared to the model used in Greenstreet et al. (2015, 2019; G15 and G19 hereafter), our model has (i) significantly fewer objects in the cold classical sub-population of the Kuiper belt, consistent with Fraser et al. (2014), and (ii) more objects in both the hot classical sub-population of the Kuiper belt and scattered disk (see Table~\ref{David-tab}, to be compared with Table I of G15 and G19). Given that the two models are based on observations and related biases (although different datasets), the origin of these differences is not fully clear to us. We suspect that this is due to our model having prior orbital distributions for each sub-population obtained from simulations of their primordial sculpting during the post-nebula migration of the giant planets.  In contrast, the G15-G19 model was built from the direct debiasing of CFEPS observations (Petit et al. 2011). Given the limited number of observed objects in this survey { (5x less than OSSOS)}, it is plausible the direct debiasing method is less reliable than a dynamics-based approach calibrated on observations, as has been shown for Near Earth Object population models (see discussion in Bottke et al., 2002).

\begin{table}[t!]
\caption{\scriptsize The { current} number of $d>100$~km KBOs in each sub-population (i.e., CCs for cold classicals, HCs for hot classicals, and SDOs for scattered-disk objects) used for our Kuiper belt model.  We use them to calibrate the number of $d>2$~km bodies that strike Pluto, Charon and Arrokoth in the current Kuiper belt per billion years as well as their mean impact velocities. We note that the total is not the sum of the reported lines as it includes also other minor KBO sub-populations not explicitly listed here { (e.g. the detached population, objects in resonances other than the Plutinos)}.}
\begin{tabular}{?c?c?m{1cm}|m{1.1cm}?m{1.1cm}|m{1.1cm}?m{1.1cm}|m{1.1cm}?}
  \hline
KBO sub-pop. & $N(d>100 {\rm km})$ & \multicolumn{6}{c|}{\# of $d>2$ km impacts per Gy $\mid$ $v_i$ (km/s)} \cr
  \hline
  &  & \multicolumn{2}{c?}{Pluto}  & \multicolumn{2}{c?}{Charon} & \multicolumn{2}{c?}{Arrokoth} \cr
\hline\hline
 CCs  & 15,000 &  \ \ 6.5&2.0  &  \ 1.3&1.8  &  0.0047&0.4 \cr
\hline
 HCs  & 40,000 & \ 46&2.6 & 10&2.4 & 0.0037&1.8\cr
\hline
 Plutinos  & 20,000 & \ 61&2.5 & 13&2.3 & 0.0022&1.2\cr
\hline
 SDOs  & 350,000 & \ 58&3.1 & 13&2.9 & 0.0045&2.4\cr
 \hline
 Total  & 550,000 & 220&2.6 & 48&2.5 & 0.018&1.5\cr
 \hline
\end{tabular}
\label{David-tab}
\end{table}

Regardless, our model yields similar impact rates to the G15-G19 model { when all impactor populations are combined together}. For Pluto, our model indicates that $5\times 10^{-11}$ bodies with $d>100$~km hit per year in the current Kuiper belt, whereas the impact rate for the same objects in G15 is $4.8\times 10^{-11}$/y. { Another estimate of the impact rate on Pluto was provided by Bierhaus and Dones (2015; BD15 hereafter): 460 impacts of $d>1$~km bodies per Gy. Given the assumed projectile SFD in that work (a cumulative power law with exponent $q_{\rm KBO}=-1.8$), this is equivalent to 132 impacts of $d>2$~km bodies, i.e. 60\% of our estimated rate.}

For Arrokoth, our model suggests that $0.018$ bodies with $d>2$~km hit per Gy, while the value estimated in G19 is 0.017.  Note that in these calculations, Arrokoth is assumed to be a 20~km diameter sphere, so the impact rate per square kilometer is $1.4\times 10^{-5}$km$^{-2}$ Gy$^{-1}$.

Given the agreement between our model and G15-G19, we call these the {\it nominal estimates} of the current impact rates. The {\it real} current impact rates are likely close to these values; { from the comparison with the results in BD15,} we expect they are within a factor of $\sim 2$. This spread corresponds to an error of only 0.2 in the slope of the SFD used to compute the number of $d>2$~km objects from the number of $d>100$~km objects; { alternatively, the factor of 2 uncertainty could be due to a smaller error on the slope, but associated with a complementary error in the estimate of the number of $d>100$~km objects.} This uncertainty in the estimate of the impact rates will be considered below.

The different balance between hot and cold sub-populations in our model and in the G15-G19 model implies some difference in the estimated mean impact velocities: they are 2.65 and 1.5 km~s$^{-1}$ for Pluto and Arrokoth, respectively, in our model while they are 2.2 and 0.7 km~s$^{-1}$, respectively, in G15 and G19. Fortunately, the crater scaling laws are only weakly dependent on impact speed (i.e., $D\propto v_i^{0.34}$ in the gravity regime; Holsapple and Housen 2007).

Assuming that the current Kuiper belt has remained steady over the last 4 Gy, the estimated impact rate on Pluto would imply $5\times 10^{-5}$ impacts of $d>2$~km projectiles per km$^2$. {According to S19 a $d=2$~km projectile would make a $D\sim 17$~km crater { on both Pluto and Charon} (see section~\ref{offset} for a discussion of projectile-to-crater scaling laws)}. On Pluto, the observed crater spatial densities on the most highly cratered surfaces are about $10^{-4}$ craters per km$^2$ for $D>17$~km (Fig. 9 in R17). This excess factor of 2 (i.e. 1--4, assuming the aforementioned uncertainty on the current impact rate relative to the nominal estimate) is also found in S19 on Charon.

The oldest surfaces in the Pluto-Charon system, however, may not be on either Pluto or Charon but rather on Nix, a 50 km diameter satellite of the Pluto-Charon system.  The spatial density of craters on Nix, once corrected for their larger size due to Nix's weaker gravity (approximately 2.1-2.6 times bigger; Weaver et al. 2016), can be compared to the spatial density of craters on Pluto. The result suggests that Nix has a spatial density of craters that is 3 times higher than the most cratered units of Pluto, with an uncertainty up to $\sim 5$ times (R17). Because Nix is unlikely to have been resurfaced by geological processes, the crater density on Nix may be the most representative case of the time-integrated real bombardment of the Pluto system. These high crater densities, suggest that the bombardment rate on the Pluto system has not been constant over the age of the Solar System. As a confirmation of this, consider the Sputnik basin on Pluto, presumably produced by a $d\sim 200$~km impactor (Johnson et al., 2016).  At the current rate, the probability of such an impact over the age of the Solar System would be only { $\sim 2.5$\%. We derived this probability using values from Table~\ref{David-tab}, where we assumed a ratio of 3,700 between the number of $d > 2$~km and $d > 100$~km objects, as found in the Trojan population, and another factor of 10 between the number of $d > 100$ and $d > 200$~km objects, as appropriate for the hot Kuiper belt. Of course, the formation of Sputnik is a singular event and therefore it has limited statistical significance; nevertheless it suggests that the bombardment rate integrated over the age of the Solar System was an order of magnitude higher than that estimated from the current rate,} which is consistent with Nix's crater record.

\subsection{Early bombardment}
\label{early}
\begin{figure}[t!]
  \centerline{\includegraphics[height=10.cm]{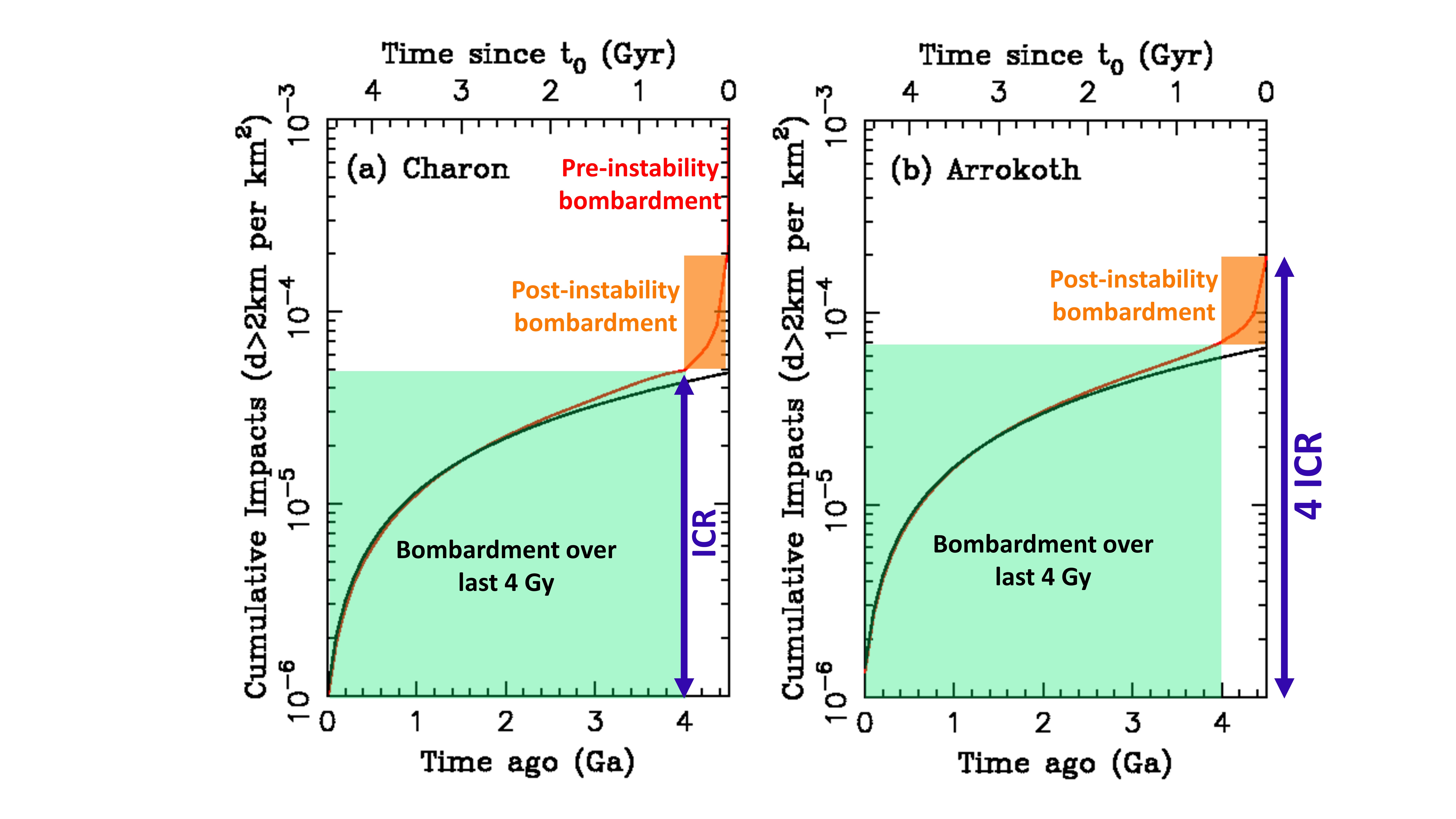}}
\caption{\scriptsize The number of impacts of projectiles with diameter $d>2$~km {per unit surface} on Charon (left panel) and Arrokoth (right panel). The black curve shows the accumulation of impacts with increasing surface age, assuming a constant bombardment at the estimated rate in the present Kuiper belt. The red curve shows instead the accumulation of impacts considering an evolving belt, from a primordial time when the trans-Neptunian disk was much more massive than now, under the migration of Neptune and the giant planet instability. Thus, the red curve deviates from the black curves for old surfaces, recording the early bombardment. { The green and orange rectangles and the red arrow highlight the bombardment cumulated during the last 4 Gy, the post-instability and pre-instability periods, respectively.} For Charon, the time-integrated bombardment since the giant planet instability should have been about 4 times the one expected from the current bombardment rate (i.e. the { top of the orange rectangle is 4x higher than the top of the green rectangle}). The bombardment before the giant planet instability is unconstrained, and we assume here it could have raised the total number of impacts by { another factor 2.5} (the vertical red segment on the vertical axis), { leading to a total enhancement of the bombardment by a factor of 10, relative to the value indicated by the top of the green rectangle}. Instead, Arrokoth, { being in the cold Kuiper belt,} experienced a strong bombardment only after the giant planet instability. {Notice that bombardment per unit surface, integrated since the time of the giant planet instability, is approximately the same for Arrkoth and Charon} (Of course, the bombardment of the two bodies has been evaluated using the same Kuiper belt evolution model). {The vertical arrow on the left of the figure defines the unit 'ICR' used in this section. The bombardment accumulated since the giant planet instability is equal to 4 ICR on both Charon and Arrokoth (right-hand side arrow).}
\label{David}}
\end{figure}

\begin{figure}[t!]
  \centerline{\includegraphics[height=10.cm]{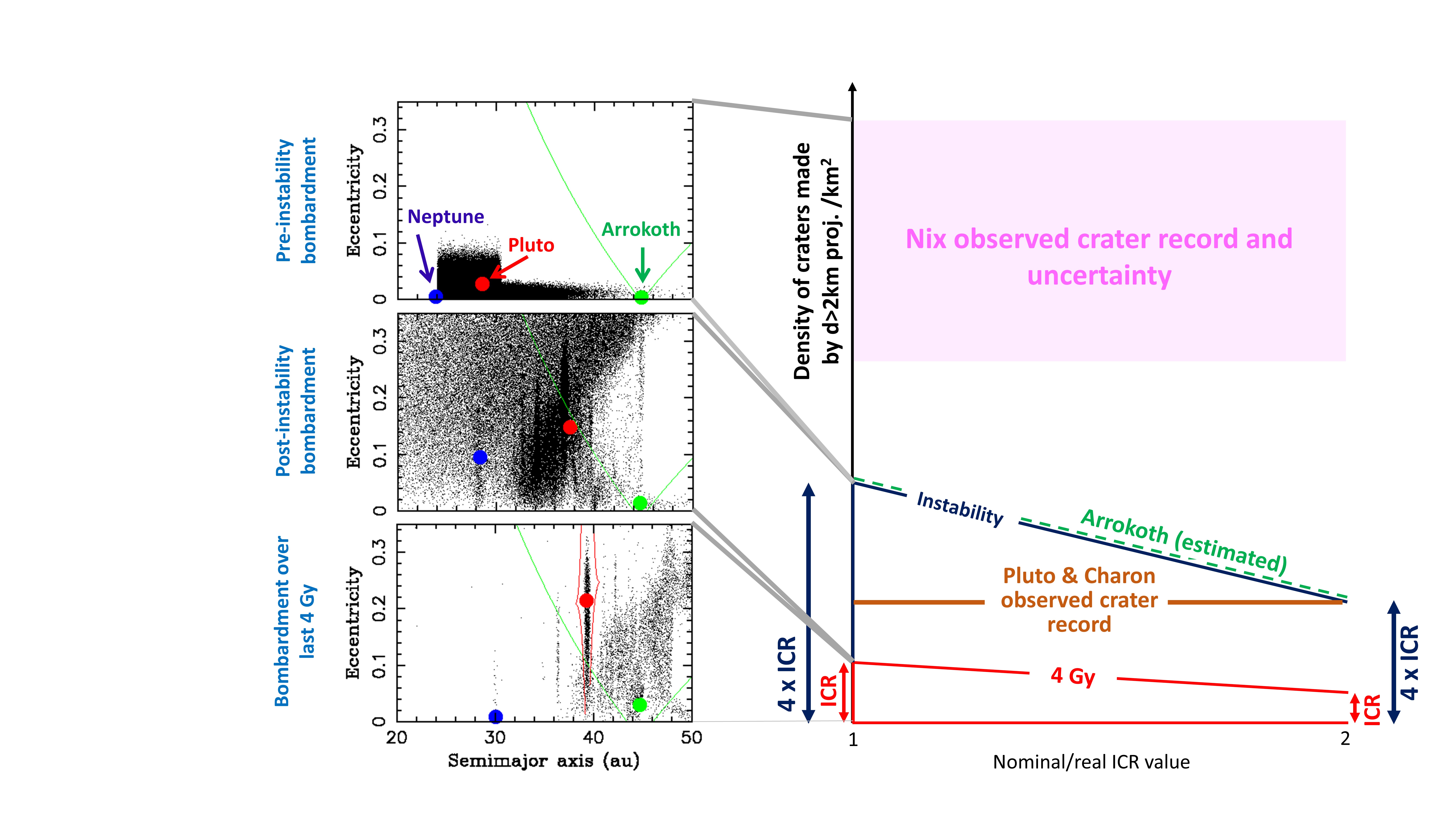}}
\caption{\scriptsize The panels on the left-hand side show the $(a,e)$ distribution of the trans-Neptunian planetesimals (black dots) in (i) the pre-instabiity phase (top) (ii) during the giant-planet instability (middle) and (iii) in the final Kuiper belt, basically unchanged since 4 Gy ago (bottom). Blue, red and greed dots denote the orbits of Neptune, Pluto and Arrokoth in these three phases. Notice that while the orbits of Neptune and Pluto change significantly, that of Arrokoth is only mildly affected. The green curves delimit the region where planetesimals cross the orbit of Arrokoth and highlight that Arrokoth could collide only with very few bodies in the pre-instability phase. The boundaries of the 2:3 resonance with Neptune are also depicted in red in the bottom panel. The diagram on the right sketches the relationship between the crater record of a body and its crater retention age. ICR is the number of impacts of projectiles with $d>2$~km per unit surface area on Pluto/Charon, cumulated over 4.5~Gy at the current impact rate. The density of impacts cumulated over the last 4~Gy is about 1 ICR. Surfaces of the Pluto system as old as the instability time would record 4 ICR. Surfaces recording the pre-instability bombardment as well, would have a crater record exceeding 4 ICR. Thus the red and blue lines labelled ''4 Gy'' and ''Instability'' mark the boundaries in this crater-retention diagram of the three phases illustrated by the panels on the left-hand side. The density of impacts on Pluto, Charon and Nix have been measured by the New {Horizons} mission (R17- see labelled brown and pink bands, whose vertical extent represent the measurement uncertainties). Unfortunately, the value of ICR is uncertain. The horizontal axis reports the ratio between our nominal estimate of ICR and its real value. This ratio is expected to be in the range 1--2. If the ratio is 1 (left axis), the surfaces of Pluto, Charon and Nix are younger; the surfaces of Pluto-Charon post-date the instability and the crater record of Nix leaves room only to a small pre-instability bombardment. If the ratio is 2 (right end of the diagram), all surfaces result older: those of Pluto and Charon are as old as the instability of the giant planets and a more substantial pre-instability bombardment is witnessed by Nix. {The green line shows the bombardment recorded by Arrokoth, namely 4 ICR (slightly shifted from the blue line for clarity). Thus, depending on the value of ICR, the number of impacts on Arrokoth per unit surface could be between 2 and 1 times that recorded by Charon.}
\label{sketchICR}}
\end{figure}

      { The previous considerations suggest that} the impact flux on the Pluto-Charon system was higher in the past. By analogy with the Moon, Mercury, Mars and the asteroids, we expect that KBOs were more heavily bombarded in the early Solar System. Our Kuiper belt model, { by construction,} provides not just the current structure of the belt but also how it evolved, and hence provides a description of this early bombardment (Fig.~\ref{David}). { This is a significant advantage with respect to the models in G15, G19 and BD15, which can only describe the current Kuiper belt. The early bombardment occurred while the Pluto system was still embedded in a primordial, massive planetesimal disk and during the post-nebula giant planet dynamical instability}. The migration and orbital excitation of Neptune  dispersed the primordial trans-Neptunian planetesimal disk and dynamically shaped both the hot classical population and the scattered disk (Nesvorny, 2015b; Nesvorny et al., 2017). For a KBO formed at the earliest times in Solar System history, our KBO evolution model leads to three bombardment phases: (i) pre-instability of the giant planets, (ii) post-instability of the giant planets, and (iii) current. { A visual representation of the dynamical state of the trans-Neptunian population during these three phases, including the orbits of Neptune, Pluto and Arrokoth, is provided by the $(a,e)$ diagrams on the left-hand side of Fig.~\ref{sketchICR}.} { The bombardment of the Pluto-system and Arrokoth during phase (iii) was described in Sect.~\ref{current}; the bombardment during the first two phases is described here. The exact timing of the giant planet instability is not known, but it should have been in the first 100~My of solar system history (Nesvorny et al., 2018). Thus we assume that the instability happened 4.5~Gy ago, i.e. 56~My after the formation of the first solar system solid minerals (the calcium alluminium-rich inclusions or CAIs). Then, this would be the date separating phase (i) from phase (ii). The transition from phase (ii) to phase (iii) is a gradual one, but we show below that the bombardment rate was basically equal to the current one since 4.0~Gy ago, which can therefore be assumed as the boundary between these phases. } 

The giant impact that formed the Pluto system requires a very low impact velocity (less than 0.7 km/s; Canup, 2005). Thus, it had to happen early in phase (i) before the self-stirring of the trans-Neptunian disk substantially increased impact speeds. It is even possible the collision took place prior to the evaporation of the Solar nebula.   With that said, { it is unknown} when the surfaces of Pluto and Charon { started} to record impact craters.  It is possible that the early surfaces had such a low viscosity that they were unable to sustain the topography of craters (see e.g. Bland et al., 2017). We also cannot rule out the possibility of global resurfacing events on both worlds. On the other hand, Nix, being a small and more distant Moon of Pluto, likely { did not experience substantial heating due to accretion, radioactive decay or tides and, consequently -like most small bodies- experienced little geological evolution. Thus, we assume that Nix should record basically the full bombardment throughout the solar system history}. Accordingly, we will follow our model backwards in time, stopping when our crater production model reproduces Nix's crater spatial densities.

Using our model, we find the integrated number of impacts on the Pluto-Charon system since { 4.5~Gy ago (i.e. during phases (ii) and (iii))} is about 4 times that { estimated} from the current impact rates on the Kuiper belt over 4.5 Gy { (i.e. the sole phase (iii))}  (Fig.~\ref{David}a). { As we said above,} the real current impact rate depends on the exact nature of the current Kuiper belt, and so it could differ from the estimated impact rate given in  Table~\ref{David-tab}, possibly up to a factor of $\sim 2$ (see Sect.~\ref{current}).  Nevertheless, because the primordial trans-Neptunian planetesimal disk scales with the current Kuiper belt, changing the estimate of the current impact rate is equivalent to changing the estimate of the $d>2$~km population in both the current Kuiper belt {\it and} the primordial disk, so the bombardment ratio between phases (ii) and (iii) would not be affected.  Our result is also insensitive to the original location of Pluto in the disk prior to it being emplaced onto its current orbit. { We verified this by computing the impacts occurring on different targets, starting at different locations of the disk and having different evolutions, but all reaching Pluto-like orbits at the end of the simulation.} 

The pre-instability phase (corresponding to the vertical red segment on the right-hand axis of Fig.~\ref{David}a), instead, depends heavily on the dynamical excitation of the disk { and the duration of the pre-instability phase (here assumed to be 56~My but significantly uncertain)}. In all cases, bombardment from this early phase is unlikely to be negligible. Depending on various assumptions, this phase could be a few times to as much as $\sim 30$ times that computed from the current Kuiper belt over 4.5 Gy, with typical values being a factor of 10.

This relationship allows us to interpret the crater records on Pluto/Charon and Nix.  We will normalize everything to the { number of impacts {on these two bodies}  expected from the current impact rate over the last 4.5~Gy, called 'Integrated Current impact Rate' (ICR). From our numerical simulations (Fig.~\ref{David}), the number of impacts over the last 4~Gy is basically 1 ICR (i.e. the top of the green rectangle is at the same level reached by the black curve at 4.5~Gy). However, a surface of the Pluto system that cumulated all the impacts received in the aftermath of the giant planet instability should have recorded 4 ICR, while a surface exceeding this value would have recorded also at least part of the pre-instability bombardment. The crater record of Pluto and Charon suggest that these bodies record 2 ICRs, which means both of their surfaces started to retain craters in the post-instability phase. Nix has 3-5 times more craters than Pluto, which would mean its crater record corresponds to 6-10 ICRs (see Fig.~\ref{sketchICR}, left axis of the right-hand diagram, for illustration). Remember that Nix's surface should be as old as Nix itself, so that it should record the bombardment since the formation of Pluto's system, expected to have occurred very early in Solar System chronology. Then Nix's crater record being equivalent to 6-10 ICRs would imply that the pre-instability bombardment was only 2 to 6 ICRs.  These values are at the low end of our range of estimates for that phase (typically 10 ICRs, possibly up to 30 ICRs, as said above).}

{ However, as we said before, we only have an approximate estimate of the current impact rate, and therefore of the value of ICR. If our nominal estimate of ICR exceeds the real value by a factor of 2, Pluto and Nix would record 4 ICRs and 12 --20 ICRs, respectively (see Fig.~\ref{sketchICR}, right side). In this case, the pre-instability bombardment would have delivered 8-16 ICRs, more in line with our theoretical expectations. Instead, if our nominal estimate of ICR underestimated the real value by a factor of 2, craters on Nix would only need 3-5 ICRs, leaving {little to} no room for any pre-instability bombardment, which is implausible. Thus, we conclude that the reality is likely bracketed within the first two possibilities (i.e., the real value of ICR is between 0.5 to 1 times our nominal estimate, which is the range reported on the {h}orizontal axis of the right-hand diagram of Fig.~\ref{sketchICR}).  We retain this spread as a measure of statistical uncertainty in the analysis that will follow.}

{ Arrokoth is a small body. Thus, like Nix, its surface should record craters since the very beginning of solar system history, when Arrokoth formed. However,} in contrast with the Pluto system, the early, pre-instability bombardment of Arrokoth was not prominent. This is because Arrokoth resides { (and presumably always resided)} in the cold sub-population, which was never substantially populated.  For this reason, its impact rate became intense only in the aftermath of the giant planet instability when the heavily populated inner disk was dispersed { (see the evolution diagrams on the left-hand side of Fig.~\ref{sketchICR})}. Thus, the model uncertainty in the pre-instability phase discussed before does not substantially affect Arrokoth's cratering history.

We find that the time-integrated bombardment on Arrokoth should have produced
{$\sim 2\times 10^{-4}$ impacts per km$^2$ with a $d>2$~km projectile for our nominal estimate of the current impact rate, corresponding  $\sim 4$ ICR. If the actual impact rate is half of our nominal estimate, the number of impacts per km$^2$ has to be divided by two but, because the value of ICR scales by the same factor, the integrated impact rate on Arrokoth would still be equal to 4 ICR. Because Pluto and Charon record  2 to 4 ICR, we conclude that the time-integrated bombardment recorded on Arrokoth { per unit surface} is 1.0--2.0 that recorded on Charon, for a given (arbitrary) projectile size. This ratio by construction already accounts for the difference in crater retention ages of the respective surfaces (the age of the solar system for Arrokoth; $\sim 4.4$--4.5~Gy for Pluto and Charon). It will be used when re-scaling the crater densities observed on the two bodies in Sect.~\ref{offset}.}

Our calculations above suggest that most of the craters on Charon and Arrokoth formed in the aftermath the giant planet instability, { i.e. in phase (ii)}.  Accordingly, to convert the impactor size into a crater size, it is appropriate to use the mean impact velocities corresponding to that phase of dispersal of the trans-Neptunian disk { rather than those corresponding to the current belt, reported in Table~\ref{David-tab}}. These velocities are significantly higher than those in the current Kuiper belt because of the presence of bodies on more eccentric and inclined orbits than those surviving today { (see left panels in Fig.~\ref{sketchICR})}. From the results of our simulations, we will adopt an impact velocity of 2~km/s on Arrokoth and 3.3~km/s on Charon.

\section{Reconciling the various constraints from Arrokoth's crater record}
\label{MonteCarlo}

As we mentioned in the Introduction, one of the most intriguing aspects of Arrokoth's crater record is the presence of a significant number of sub-km craters and the absence of craters with $D>1$~km except for Maryland crater ($D\sim 7$~km).

The crater database of Sp20 reports 5 craters with $0.69<D<0.96$~km identified with high confidence. Given the error bars reported on the sizes of the individual craters, we will assume that these craters fill the size-range 0.55--1.15~km. The surface of Arrokoth visible during the New Horizons flyby was 700~km$^2$. Although craters have been searched across the visible surface, it is clear from Fig. 6 in Sp20 that the craters identified with high confidence are located on a much smaller region where the view geometry or the illumination conditions were particularly favorable (e.g., on the small lobe, on the rim of the Maryland crater or near the terminator). It is difficult to precisely determine the effective area on which craters could be detected with confidence. From a visual inspection of Figs. 1a and 6 in Sp20, we estimate the area may only be $\sim 25$-50\% of the visible surface. Thus, the spatial density of craters in the considered size-range would be 0.014--0.028 per km$^{2}$. This value agrees well with that estimated for the craters detected on the sole terminator (2 in 90~km$^2$, yielding a spatial density of 0.022 per km$^2$) or for the pit-shaped craters (3 in 230~km$^2$, for a spatial density of 0.013 per km$^2$). { Accordingly, below we will consider the density of 0.55--1.15~km craters to be $0.021\pm 0.007$ per km$^2$.}

\begin{figure}[t!]
  \centerline{\includegraphics[height=10.cm]{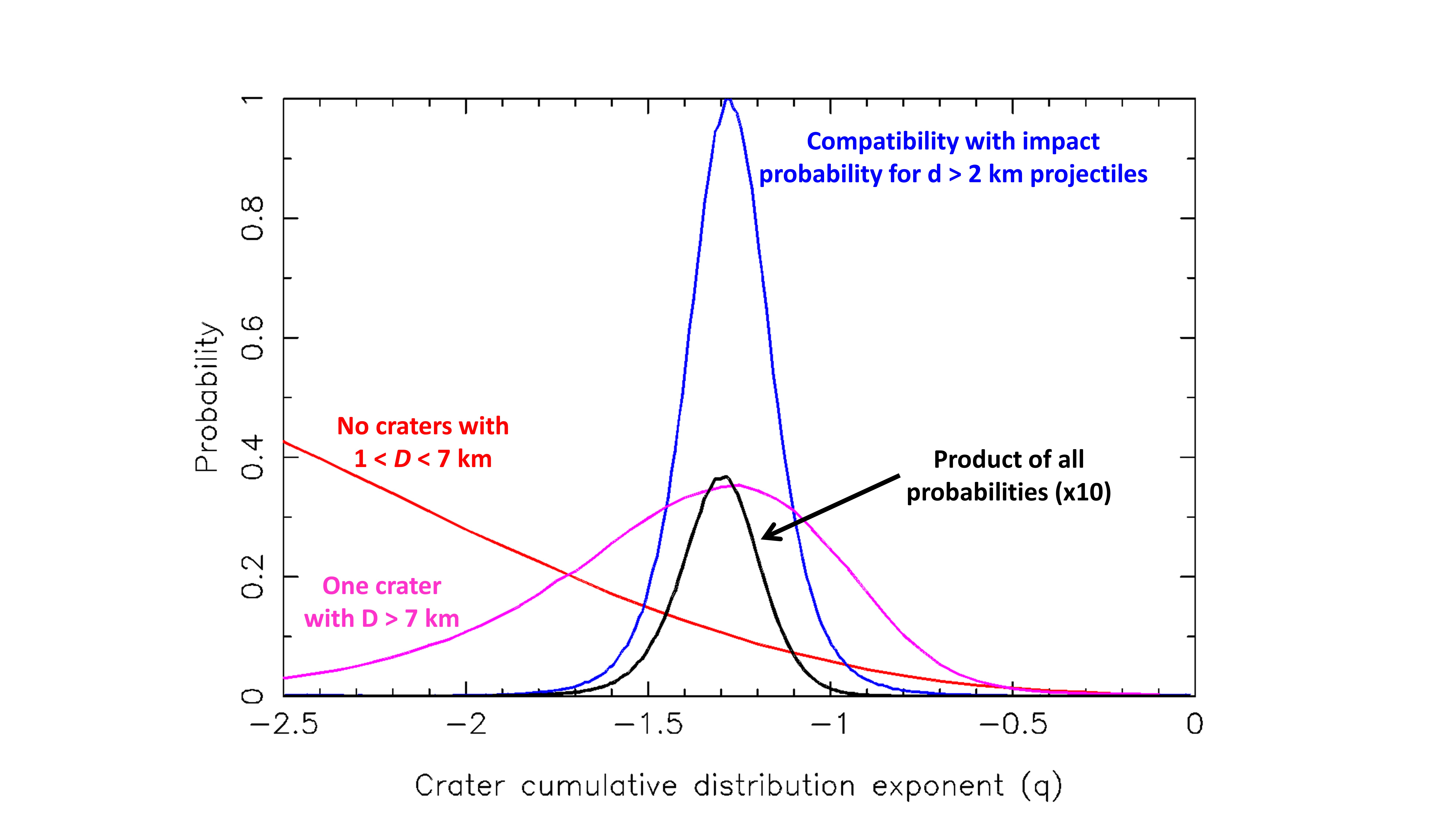}}
\caption{\scriptsize This diagram shows, as a function of the cumulative slope of the SFD of the crater-production function, the probability that one of several constraints related to Arrkoth's crater record is satisfied. The red curve shows the probability that no crater with $1.15<D<7$~km formed in the same region where craters with $0.55<D<1.15$~km are observed. The magenta curve depicts the probability that one and only one Maryland-like crater (i.e. $D>7$~km) formed on the visible surface. The dark-blue curve is the probability that the density of $0.55<D<1.15$~km craters is consistent with the estimated formation probability of a $D>27$~km transient crater (i.e. the impact probability of a $d>2$~km projectile). Finally, the black curve is the product of the red, light-blue and dark-blue curve, scaled up by a factor of 10 for readability.
\label{probas}}
\end{figure}

Given five craters in the considered size-interval and a slope $q$ of the cumulative crater SFD, we used a Monte Carlo simulation to compute the probability that no craters with $1.15<D<7$~km formed on the { \it same} surface where the aforementioned sub-km craters were observed. { Note that this} calculation is independent of the surface area used { (because it is the same for the two considered crater size-ranges), which is important given that the effective surface area where craters could be visible is} uncertain, as discussed above. The result is illustrated as a function of $q$ by the red curve in Fig.~\ref{probas}. The probability is relatively high for steep SFD ($\sim 40$\% for $q=-2.5$) and decays monotonically with increasing $q$. For $q=-0.7$ (the value reported in S19 { for craters on Charon}) the probability is only 2.2\%, { but for $q=-1.3$ (the value reported in Sp20 for craters on Arrokoth), the probability is $\sim 10$\%.}

%On the other hand, large craters should have been visible on a much broader surface than sub-km craters. For example, as a thought experiment, let us assume that a crater with $D>2.5$~km would be visible on the entire imaged surface of Arrokoth. In our Monte Carlo simulation, we can compute the probability that no crater with $2.5<D<7$~km formed on the full 700~km$^{2}$ visible surface while also applying the constraint that the spatial density of 0.55--1.15~km craters has a Gaussian probability distribution of $0.02\pm 0.008$ per km$^2$. The result is illustrated as a function of $q$ by the magenta dashed curve in Fig.~\ref{probas}.  The probability of getting a steep slope, such as $q\lesssim -1.5$, is now larger than before, while it is comparable to the previously computed probability for shallow SFDs. Of course, this probability would increase/decrease with the increasing/decreasing size of the craters assumed to be detectable on the full surface. Thus, this probability is indicative but not particularly constraining and we will not consider it any further in our analysis.

{ Another important} scenario to consider is whether a projectile SFD could plausibly produce a single $D>7$~km crater on the visible surface of Arrokoth as well as the spatial density of $0.021\pm 0.007$ per km$^2$ for 0.55--1.15~km craters. Our Monte Carlo results are illustrated by the magenta curve in Fig.~\ref{probas}. The probability of occurrence is very low for steep SFDs but it increases with increasing $q$ until a maximum probability is reached for $q\sim -1.25$. The probability then declines again for shallower slopes. This is due to the fact that very shallow slopes of the crater-production function are likely to produce more than one big crater.  We note that the very shallow power law slope estimated { for Charon's craters} in S19, $q=-0.7$, only yields a small probability of occurrence (5.4\%). { Instead, the maximum of this probability function is very close to the slope $q=-1.3$ found in Sp20 for Arrokoth's craters. The agreement between these two results and Sp20 should not surprise because our analysis so far is equivalent to that done in that paper.}

Another important constraint, { not considered in Sp20,} is provided by the link between the spatial density of sub-km craters on Arrokoth and the probability of an impact of a $d>2$~km projectile over the age of the Solar System, estimated in the last section to be $1.5\pm 0.5 \times 10^{-4}$ per km$^2$ {(accounting for the uncertainty on the value of ICR, which could be smaller by a factor of 2 than the one reported on Fig.~1)}. Assuming an impact velocity of 2~km/s, a projectile with the same density as Arrokoth, a surface gravitational acceleration of $0.5$mm/s$^2$ (Keane et al., 2020) and a scaling law in the gravity regime for a porous sand-pile target (Holsapple and Housen, 2007; see (\ref{SLg}) below), we find that a $d=2$~km projectile would form a $D=27$~km transient crater on Arrokoth\footnote{ This crater is larger than Arrokoth itself, meaning that the collision of a $d=2$~km projectile would be catastrophic for the target. However, for the purpose of defining a crater-production SFD, proxy of the projectile SFD, it is legitimate to consider this virtual crater-formation event. {Notice that, integrated on the surface of Arrokoth, the probability of such a collision is less than 0.25, which is consistent with the fact that this event never happened.}}. Given the very low gravity of Arrokoth, we neglect crater collapse, and consider the transient crater as final crater. Thus, the cumulative power law slope of the crater SFD has to be such that a spatial density of  $0.021\pm 0.007$ per km$^2$ for craters with $D>0.55$~km is consistent with a formation probability of $1.5\pm 0.5 \times 10^{-4}$~km$^{-2}$ for a crater with $D>27$~km. In other words, we have two points, and we want to use them to estimate the slope of the line connecting them on a log-log plot.  

{ Assuming that the reported uncertainties are Gaussian and} using a Monte Carlo code, we obtain a probability distribution for the cumulative slope of the SFD which is illustrated by the dark-blue Gaussian in  Fig.~\ref{probas}. This distribution peaks at $q\sim -1.3$, { again the slope of Sp20,} with a width $\sigma=0.1$. Note that the probability for $q=-0.7$ is only 0.4\% because such a shallow SFD would predict too great a likelihood of $d>2$~km impacts compared to our estimates in the previous section.

If we now multiply the probabilities illustrated by the red, light-blue and dark-blue curves in Fig.~\ref{probas}, we obtain the combined probability distribution.  It is depicted by the black curve after being magnified by a factor of 10. This probability distribution has approximately the same Gaussian shape as the dark-blue curve, i.e. constraining $q=-1.3\pm 0.1$. %The peak value of this function is not very high, only 3.7\%, but that is because the three probabilities have been multiplied together. The absence of craters larger than 1~km but smaller than Maryland is clearly the most penalizing constraint. Consider that if a single crater larger than this limit had passed unnoticed, this constraint would be relaxed, with the probabilities favoring $q=-1.3$ increasing from 13\% to 34\%.

Note that, although our preferred slope is the same as in Sp20, the error bar on the slope determination is much smaller in our work (0.1 instead of 0.6 in Sp20), thanks to the constraint provided by the last analysis, illustrated by the dark-blue curve in Fig.~\ref{probas}. {Consequently, whereas Sp20 found that the slopes of the crater SFDs on Arrokoth and Charon were consistent with each other within error bars, we can rule out that the slopes reported for Charon in S19 and Sp20 apply to Arrokoth, with a confidence of 6~$\sigma$ and 5~$\sigma$, respectively.}

  One might argue that the disagreement is only apparent, because Arrokoth and Charon may be probing different projectile populations or different projectile size-ranges. {We showed in Sect.~\ref{ImpRates} that most craters on Charon and Arrokoth come from phase (ii), i.e. in the aftermath of the giant planet instability, so that they should have been caused by the {\it same} population of impactors. However, the impact velocities on the two bodies are different (2~km/s and 3.3~km/s for Arrokoth and Charon respectively, see Sect.~2) and different projectile-to-crater scaling laws may apply on the two worlds. Thus, we compare in the next section the absolute crater densities measured on the two worlds, accounting for the appropriate scalings. This will allow us also to check the statement in Sp20 that Arrokoth seems to be more heavily cratered than expected from the extrapolation of Charon's crater SFD to sub-km craters.}

\section{Rescaling Arrokoth's to Charon's crater densities}
\label{offset}

{We consider here the spatial density of km-size craters and of Maryland, measured on Arrokoth. The first step to rescale Arrokoth's crater record to Charon's is to compute the sizes of the craters formed on Charon by the impact of projectiles of the same size as those responsible for the aforementioned craters on Arrokoth.}

The scaling law relating projectile size { $d$} to crater size { $D$} in the gravity dominated regime is, from Holsapple and Housen (2007):
\begin{equation}
  D = c_g d \left({{g d}\over{2 U^2}}\right)^{K_g} \left({{\rho_i}\over{\rho_t}}\right)^{1/3}\ ,
  \label{SLg}
\end{equation}
where $g$ is the surface  gravity of the target, $U$ is the vertical impact speed (on average equal to $v_i/\sqrt{2}$, where $v_i$ is the impact velocity) and $\rho_i$ and $\rho_t$ are the impactor and target densities, respectively. The units have to be chosen so that the quantities in each parenthesis are dimensionless. The coefficients $c_g$ and $K_g$ are 1.03 and $-0.17$ if the target soil is dominated by regolith and 1.17 and $-0.22$ if the target soil is non-porous (e.g. ice), { respectively}.

The scaling law (\ref{SLg}) can produce non-physically large craters on small bodies where $g$ is very small. In this case, the projectile-to-crater scaling law is in the strength regime, and is (again from Holsapple and Housen, 2007):
\begin{equation}
  D = c_s d \left({{Y}\over{\rho_t U^2}}\right)^{K_s} \left({{\rho_i}\over{\rho_t}}\right)^{2/5}\ ,
  \label{SLs}
\end{equation}
where $c_s=1.03$ and $K_s=-0.205$ (again for dusty regolith) and $Y$ is the material strength (again units should be chosen so that the quantities in each parenthesis are dimensionless).

Because Arrokoth is a porous, low-density body (its bulk density is $0.24$g/cm$^3$, with a $1\sigma$ range from 0.16 to 0.45; Keane et al., 2020), we first adopt a scaling law (\ref{SLg}) with $c_g=1.03$ and $K_g=-0.17$. We assume again that the equatorial surface gravity of Arrokoth is $g=0.5$~mm/s$^2$. For an impact velocity of 2~km/s (see the end of section 2) and assuming the projectile has the same density as Arrokoth itself, we find that a km-size crater would have been produced by a 37~m projectile. {Instead, a crater like Maryland ($D=7$~km) could be formed by a $D=390$~m projectile. } 

As a comparison, we have also repeated the calculation in the strength regime. Unfortunately, the strength of Arrokoth's material is unknown. The strength of the material of comet 67P/C-G at a scale of a 1--100~m is a few Pascal (Attree et al., 2018). For such a low value, km-size cratering would definitely occur in the gravity regime on Arrokoth. However, Arrokoth is significantly larger than 67P/C-G, potentially implying a more compact harder soil. If we assume that the strength of Arrokoth is 500~Pa (the strength of snow; Sommerfeld, 1974), then the cratering process occurs in the strength regime and, using (\ref{SLs}), we find that the projectile size required to make a km-sized crater is $d=57$~m.  This value is comparable to the projectile's size computed in the gravity regime. Thus, we will consider the range of possible projectile sizes for Arrokoth's km-sized craters to be 37 and 57~m. {For Maryland, given the higher impact energy, the use of the scaling law in the gravity regime is more appropriate, but using the strength regime would yield basically the same result: $D=400$~m.}

Charon has a much larger gravity than Arrokoth ($g=0.278$~m/s$^2$), so its craters definitely occur in the gravity regime.  A big unknown for Charon craters, however, is the nature of the surface layer that would be excavated by a sub-km crater. This makes it unclear whether the scaling laws for non-porous or regolith material should be chosen. In case of non-porous material, a reasonable density for the surface material would be 0.75--1.0~g/cm$^3$, namely $\sim$1.5 to 6 times that of Arrokoth. In the case of regolith, a density of 0.5--0.75~g/cm$^3$ is plausible, corresponding to $\sim$1--4 times the density of Arrokoth. Table~\ref{Charon-crat} reports the crater sizes on Charon for all these possibilities using the projectile sizes {determined above} that can make a $D_{(A)}=1$~km or a $D_{(A)}=7$~km crater on Arrokoth, {as well as for a $d=2$~km projectile}. The impact velocity on Charon is assumed to be 3.3~km/s (see end of Section~2). 

\begin{table}[t!]
\caption{\scriptsize Crater size $D_{(C)}$ on Charon for the considered projectile sizes on Arrokoth. {For a $d=2$~km projectile, we only consider a non-porous Charon, because the crater would presumably excavate deeper than the regolith layer; in this case, the $\rho_C/\rho_A$ ratio has to be intended as the target/projectile density ratio.}}
\begin{tabular}{|c|c|c|c|c|}
\hline
Charon's soil   & $d=37$~m  & $d=57$~m & $d=390$~m & $d=2$~km\cr
\hline\hline
non-porous, $\rho_C=1.5\rho_A$   & $D_{(C)}=$0.81~km & $D_{(C)}=$1.13~km & $D_{(C)}=$5.0~km & $D_{(C)}=$18.1~km\cr
\hline
non-porous, $\rho_C=6\rho_A$   & $D_{(C)}=$0.53~km & $D_{(C)}=$0.78~km & $D_{(C)}=$3.3~km & $D_{(C)}=$11.8~km\cr
\hline
porous, $\rho_C=\rho_A$   & $D_{(C)}=$0.40~km & $D_{(C)}=$0.58~km & $D_{(C)}=$2.8~km & \cr
\hline
porous, $\rho_C=4\rho_A$   & $D_{(C)}=$0.26~km & $D_{(C)}=$0.36~km & $D_{(C)}=$1.8~km & \cr
\hline
\end{tabular}
\label{Charon-crat}
\end{table}

For reference, had we used impact velocities corresponding to the current Kuiper belt, rather than those appropriate for the predominant early bombardment, i.e. 2.3~km/s and 1.1~km/s for Charon and Arrokoth respectively (taking the average between the values in our model and G15-G19), the projectile sizes would have been bigger ({for instance} 48 and 73~m {for the formation of km-size craters on Arrokoth} in the gravity and strength regime), but the ratio between the sizes of the craters on Charon and Arrokoth would have changed very little (because the impact velocities on both bodies are proportionally reduced), ranging from a minimum of 0.27 to a maximum of 1.37. Thus, within the broad uncertainty reported in Table~\ref{Charon-crat}, the results are robust.

\begin{figure}[t!]
\centerline{\includegraphics[height=10.cm]{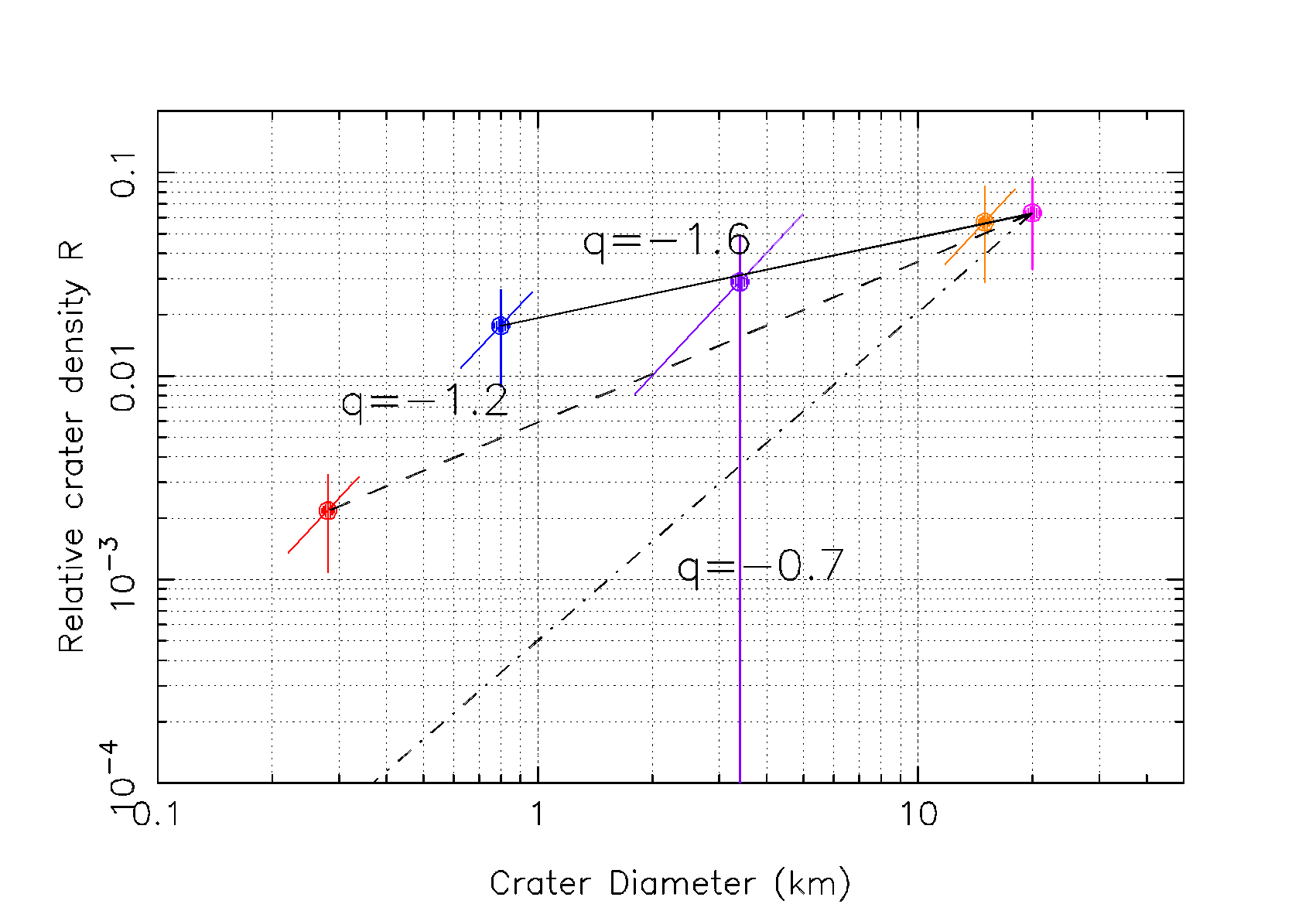}}
\caption{\scriptsize $R$-plot summarizing the crater density measured on Charon at $D_{(C)}=20$~km (magenta dot with vertical error bar) or rescaled from {Arrokoth for $D_{(A)}=0.85$~km craters (red and blue dots), Maryland (violet dot) and the virtual $D_{(A)}=27$~km crater (orange dot).}  The red dot has been obtained by assuming that $D_{(C)}/D_{(A)}=0.33$. The blue dot was obtained by assuming that $D_{(C)}/D_{(A)}=0.94$. These ratios bracket the possibilities illustrated in Table~\ref{Charon-crat}. The slanted error bars represents the error due to the uncertainty on {the size of the rescaled crater on Charon}. The vertical error bar reflects the uncertainty on the measured crater spatial density on Arrokoth and on the ratio of impacts suffered by the two bodies over the age of their respective surfaces. The solid and dashed black lines show the resulting exponents of the cumulative crater size distributions implied by the blue and red datapoint, respectively, {together with the magenta dot}. The uncertainties on the crater scaling law to be adopted for {Charon's small craters} essentially allow any slope within these two lines. The dash-dotted line shows the slope in the $R$-plot for a cumulative crater size distribution of $q=-0.7$, as { measured in S19 for Charon}. 
\label{results}}
\end{figure}

{ Geologists working on crater counts usually do not consider the incremental number of craters per unit area ${\delta N(D,\delta D)}$, where $D$ is the center of the considered bin of crater sizes of width $\delta D$, but another quantity, called $R$ defined as $R=\delta N(D,\delta D) \times D^3 / \delta D$. This is done to highlight the difference between the measured SFD from a cumulative power law with exponent $q=-2$. On Arrokoth, the density of craters $\delta N=0.021\pm 0.007$ was measured for a crater-size range of 0.69--0.96~km, if the nominal crater sizes are considered, or 0.55-1.15~km, if the uncertainty on the sizes is also included (Sect.~\ref{MonteCarlo}. Accordingly, we assumed $\delta D=0.45$~km (the average of these two intervals) and $\bar{D}=0.85$~km. This gives a conversion factor between ${\delta N}$ and $R$ of 1.5 for the considered size-bin.  Thus, the $R$-value for Arrokoth, hereafter denoted $R_{(A)}$, for $D=0.85$~km is $0.03\pm 0.010$.  This value is consistent with the broad uncertainty reported in Fig. 6 of Sp20.}

{ With the values reported in Table~\ref{Charon-crat}, we now have all the elements to convert $R_{(A)}$ to the equivalent $R$-value for Charon, hereafter denoted $R_{(C)}$.}  Our method is as follows. First, $R_{(A)}$ has to be divided by the impact ratio Arrokoth vs. Charon for a given projectile size, integrated over the bombardment recorded on the respective surfaces  (i.e., 1.0--2.0; see Section~2). Second, because we are considering crater spatial densities in a $R$-plot, $R_{(A)}$ has to also be multiplied by $[D_{(C)}/D_{(A)}]^2$, { where $D_{(A)}$ is the considered crater size on Arrokoth (here 0.85~km) and $D_{(C)}$ is the size of the crater that the same projectile would cause on Charon.}

Fig.~\ref{results} shows the resulting $R_{(C)}$-values for Charon for $D_{(C)}/D_{(A)}=0.33\pm 0.07$~km (red filled dot) and $D_{(C)}/D_{(A)}=0.94\pm 0.20$~km (blue filled dot).  They correspond to the averages and deviations of the values in the last two rows in the second column and to the first two rows in the third column of Table~\ref{Charon-crat}.  The error bar for $D_{(C)}$ in Fig.~\ref{results} is inclined with a slope of 2 to reflect the dependence of $R_{(C)}$ on $D_{(C)}$. The vertical error bar comes from combining the uncertainty of the measurement of $R_{(A)}$ on Arrokoth ($\pm 30$\%) and the uncertainty on the impact ratio between Arrokoth and Charon (also $\pm 30$\%).  It amounts to $\pm 42$\% of $R_{(C)}$.

{We proceed in a similar way for Maryland and the 27~km virtual crater produced by a $d=2$~km impactor on Arrokoth. The value $R_{(A)}$ for Maryland is $0.18^{+0.12}_{-0.18}$ (Sp20). From the average of the values reported in the 4th column of Table~\ref{Charon-crat} the size of the crater produced on Charon by the projectile that made Maryland is $D_{(C)}=3.4\pm 1.6$~km. Thus, the corresponding $R$-value for Charon is $R_{(C)}=R_{(A)}/(1.5\pm 0.5) [(3.4\pm 1.6)/7]^2$. This is depicted by the violet filled dot and its error bars on Fig.~\ref{results}. With a surface density of $1.5\pm 0.5 \times 10^{-4}$~km$^{-2}$, the $R$-value for the virtual 27~km crater on Arrokoth is $R_{(A)}=0.28\pm 0.09$. From the last column of  Table~\ref{Charon-crat} the size of the crater produced on Charon by a $d=2$~km projectile is $D_{(C)}=15\pm 3$~km. Thus, the corresponding $R$-value for Charon is $R_{(C)}=R_{(A)}/(1.5\pm 0.5) [(15\pm 3)/27]^2$. This is depicted by the orange filled dot and its error bars on Fig.~\ref{results}.}

{Next, we consider the $R$-value for $20$~km craters ($R_{(C)}(20 {\rm km})=0.063\pm0.03$; magenta dot in Fig.~\ref{results}), which comes from the average of the values measured directly on Charon for craters of 15, 20 and 30~km and their uncertainties (Fig. 6 of Sp20). We remark that the fact that the orange dot, scaled from Arrokoth, falls so close to the magenta dot, directly measured on Charon, is a relevant validation of all the scalings implemented in this section.}  

{To measure the slope of the crater production SFD and its uncertainty, we connect the red and blue dots with the magenta dot.}  The resulting slopes on the $R$-plot are $0.78\pm 0.17$ and $0.38\pm 0.23$ (plotted on Fig.~\ref{results} as dashed and solid lines).  They translate into the exponents of the cumulative power law slope of the crater production SFD with $q=-1.22\pm 0.17$ and $q=-1.62\pm 0.23$, respectively. By the nature of our approach, any slope between these two values is possible and there is a legitimate projectile-to-crater size relationship that can produce it. {Note that all these slopes pass through the error bars of the violet and orange points, so they are all equally acceptable}. 

%We find it intriguing that $-1.2 < q < -1.6$ encompasses the interval constrained in Section~3 (see Fig.~\ref{probas}), although it is a bit larger and its center is slightly shifted towards a steeper slope ($q=-1.4$ vs. $q=-1.3$). Given the uncertainties in all these analyses, we consider this match to be satisfactory, particularly given that the approaches followed in Section 3 and in this section are independent of one another. 

{Because the orange dot and the magenta dot are so close to each other, the results of this section are very similar to those of the previous section. By combining the results of the two sections, we conclude that the power law slope for the crater production SFD on Arrokoth is $-1.2 < q < -1.5$. This range of slopes is consistent with that measured for Arrokoth alone ($q = -1.3 \pm 0.6$; Sp20), the one reported in R17 for Charon ($q = -1.47 \pm 0.15$), and is within the uncertainty range for that reported in Sp20 for Charon ($q = -0.8^{+0.4}_{-0.6}$), but is incompatible with the shallower slope argued in S19 ($q = -0.7 \pm 0.3$, dash-dotted line in Fig.~\ref{results}; see also Fig.~\ref{probas}). We thus conclude that the power law slope for the crater production SFD consistent with
both bodies is $-1.2 < q < -1.5$}.

\section{Conclusions and Discussion}
\label{conclusions}

Craters imaged on Arrokoth by the New Horizons mission have made it possible to investigate the KBO population at sizes as small as several tens of meters in diameter, i.e. much smaller than the projectiles responsible for the formation of the craters visible on Pluto and Charon. Using these data, we have examined the hypothesis formulated after the Pluto/Charon flyby that the KBO population is so shallow that it is unlikely to be in collisional equilibrium for sub-km KBOs (S19).  A brief summary of our procedure to constrain the power law slope of small KBOs is given below.  
\begin {itemize}
\item We used a model of the current Kuiper belt population calibrated on the observations from the OSSOS survey (Nesvorny 2015a,b; Nesvorny et al., 2017) that also describes the enhanced bombardment rate characterizing the early Solar System. We further calibrated this model using the crater record on Pluto, Charon and Nix. With this model, we estimated the probability of impacts from $d>2$~km bodies on Arrokoth throughout the history of the Solar System and its ratio relative to Charon (Section 2). 
\item Arrokoth's observed terrain shows many sub-km craters and a single major crater with $D\sim 7$~km. No craters were observed between these sizes. We ran Monte Carlo simulations to determine the power law slopes of the crater production SFD that are statistically compatible with these observations and with the estimated impact probability of $d>2$~km KBOs on Arrokoth (Section 3).
\item Using the estimated  ratio of impacts on Arrokoth and Charon as well as projectile-to-crater scaling laws appropriate for these two bodies, we { combined the crater records of the two objects and constrained once again the power law slope of the crater SFD.} 
\end {itemize} 
{ While each of our different lines of analysis above have uncertainties, we find it interesting and potentially compelling that they all converge to a power law slope of the crater production SFD between -1.5 and -1.2.}  This range not only matches the power law slope determined independently on Arrokoth's crater record alone (Sp20), { but it is significantly steeper than the values determined from Charon alone (i.e. $-0.7 \pm 0.3$ in S19 and $-0.8^{+0.4}_{-0.6}$ in Sp20).} It is more in line with an earlier estimate from R17, though we cannot say whether that similarity is fortuitous. { The discrepancy with S19 and Sp20 may suggest that the Charon crater database used in those works is incomplete at small sizes. On the other hand, we remark that our result is within the uncertainty reported by Sp20 for Charon (where the acceptable values of $q$ for Charon range down to $-1.4$). Thus, most likely it is just the uncertainty on $q$ reported in S19 which was poorly underestimated.} 

The power law slope of the projectile (KBO) size distribution can be calculated from these results, but we must first consider the effect of the crater scaling law applied in this circumstance.  Generically, if the crater size distribution has a differential form $N(D){\rm d}D=D^{q-1}{\rm d}D$, and the crater-to-projectile scaling law is of type $D\propto d^\alpha$, the projectile differential distribution is $N(d){\rm d}d=d^{q\alpha-1}{\rm d}d$. This translates the exponent of the cumulative crater distribution $q$ into the exponent $q\alpha$ for the cumulative projectile distribution. 

In the calculations presented in this work we have dealt with simple craters.  Thus the appropriate value of $\alpha$ is 0.78--0.83, depending on the nature of Charon's soil (Holsapple and Housen, 2007). From the analysis in Section~4, the former value is more appropriate for the steeper estimate of the crater SFD and the latter for the shallower estimate. Thus, the power law slope range of the crater production SFD and the cumulative KBO SFD goes from $-1.5<q<-1.2$ to $-1.2 < q_{\rm KBO}<-1.0$, respectively.

These slopes are quite shallow, validating the general conclusions of S19.  With that said, they are not so shallow as to rule out the possibility that the KBO size distribution is in collisional equilibrium. Consider that the main asteroid belt, which is in collisional equilibrium (e.g., Bottke et al. 2015), has a power law slope of the cumulative size distribution $q_{\rm MB}=-1.29\pm 0.02$ for bodies 0.6 and 1 km in diameter (Yoshida and Nakamura, 2007).  A similar result of -1.3 was found by Heinze et al. (2019). Both values are very close to the one determined in this work for KBOs.

Curiously, while the asteroid and Kuiper belt SFDs have similar slopes for objects ranging from hundreds of meters to a few kilometers in diameter, they diverge for objects smaller than 200~m.  For the asteroid belt SFD, the slope steepens up to approximately $q = -2.7$ for $d < 200$~m (e.g., Heinze et al. 2019; Bottke et al. 2015), with the slope controlled by the shape of the asteroid disruption law in this size range (O'Brien and Greenberg 2003).  The power law slope for the KBO SFD, however, appears to maintain $-1.2 < q_{\rm KBO}<-1.0$ all the way down to $\sim 30$~m bodies (i.e., between $0.03 < d < 1$-2~km). { The size of $\sim 30$~m comes from the fact that the shallow crater SFD slope of $\sim -1.3$ holds for craters on Arrokoth as small as 0.5--1~km (Sp20) and the projectile-to-crater conversion for Arrokoth computed in Sect.~\ref{offset}.}

This difference { in the lower-end size of the shallow SFD portion} between KBOs and asteroids is plausible, because the size at which a SFD is expected to upturn is the one below which fragmentation occurs in the strength regime (O'Brien and Greenberg, 2003); if small KBOs really have strengths of a few Pascal, as measured for 67P/C-G (Attree et al., 2018), it makes sense that the SFD of the KBO population upturns at very small sizes, possibly even below a few tens of meters. Nevertheless, { to support a claim of collisional equilibrium, we need to find direct evidence that the SFD steepens up in a fashion similar to that of the asteroid belt at small sizes.}

{ We claim that the evidence comes from the abundance of} Kuiper belt dust, which is derived from collisional processes. Poppe et al. (2019) estimate that the mass of particles between 0.5 and 500$\mu$m derived from the Kuiper belt is $3.5 \times 10^{18}$~kg.  For reference, this is the mass equivalent of $5.8 \times 10^{-7}$ Earth masses or a single 190 km diameter body made of ice (bulk density of 1000 kg~m$^{-3}$). Studies of interplanetary dust particles (Grun et al. 1985; Brownlee and Love, 1993) indicate the majority of the mass in this size range is in the form of 200$\mu$m particles.  Using this value, and a density of 1 g/cm$^3$, we estimate there are the order of $8\times 10^{26}$ such particles.

\begin{figure}[t!]
\centerline{\includegraphics[height=10.cm]{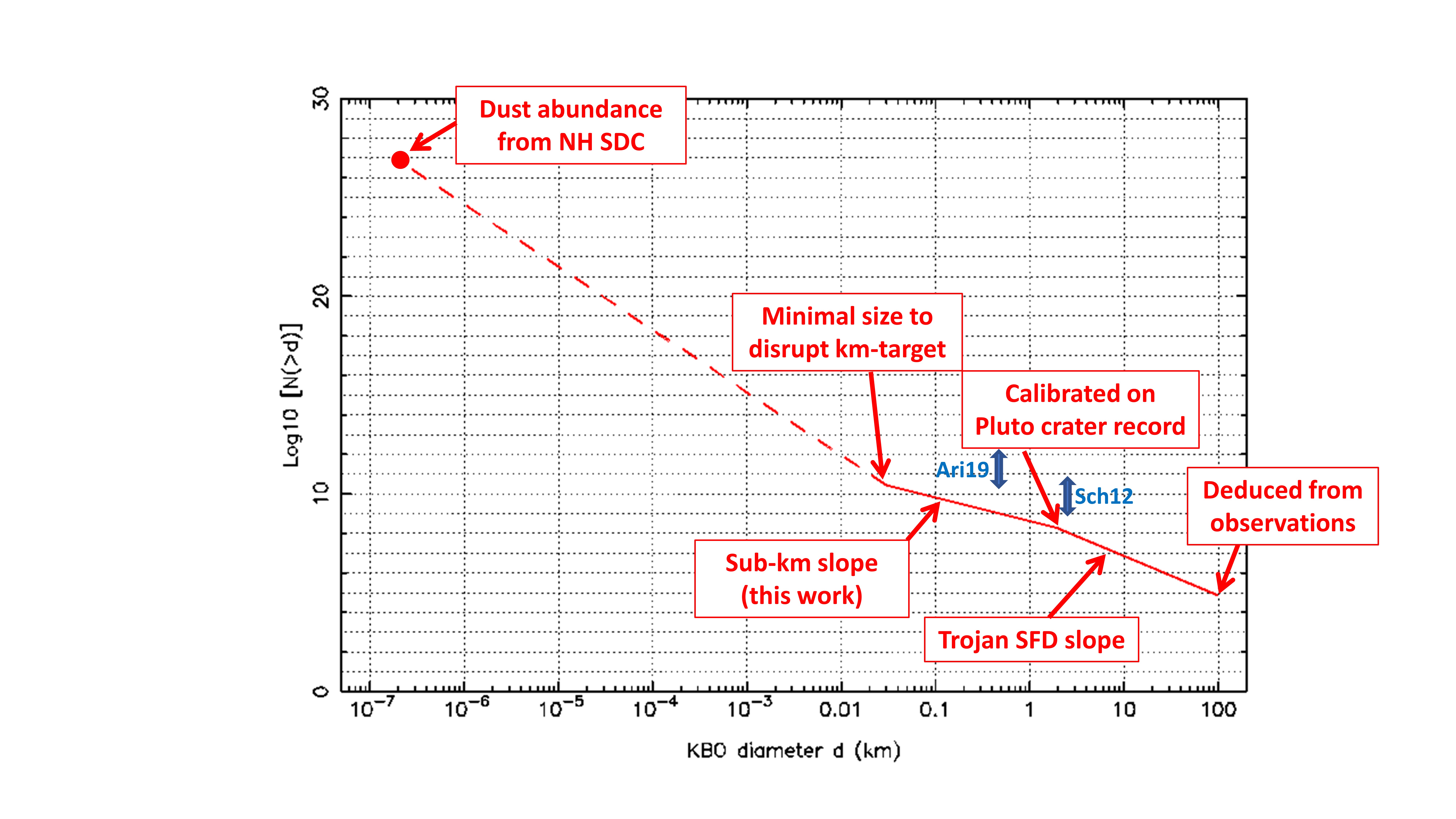}}
\caption{\scriptsize A visual summary of the various constraints on the SFD of the KBO population. The number of $d>100$~km bodies comes directly from telescope surveys. The slope for $2<d<100<$~km comes from telescope observations of KBOs and of Jupiter's Trojans (derived from KBOs). The number of $d>2$~km objects has also been refined in Sect.~2 by calibration on the crater records in Pluto's system. The slope for $0.03<d<2$~km ($q_{\rm KBO}=-1.2$) is in the range of those determined in Sections~3 and~4. The number of $200 \mu$m dust particles is { derived} from the results of the Student Dust Counter onboard New Horizons. { For reference, we also report with blue arrows the number of bodies with $d>2.5$~km and $d>0.5$~km estimated by Schlichting et al. (2012) [labelled ``Sch12''] and Arimatsu et al. (2019) [labelled ``Ari19''] from statistics of claimed stellar occultations. The slope between these two datapoints is consistent with the SFD determined in this work, but there is an unexplained vertical offset in absolute numbers.}
\label{fullSFD}}
\end{figure}

This constraint can be compared with an estimate of the inferred Kuiper belt SFD (Fig.~\ref{fullSFD}).  { Observations suggest that there $\sim$75,000 bodies with $d \ge 100$~km  in the classic Kuiper belt (cold and hot together; {Table~\ref{David-tab}}); assuming a slope for the cumulative SFD of $-2.1$, as for Jupiter Trojans, there are $2.8 \times 10^{8}$ bodies with $d \ge 2$~km}.  If the cumulative SFD between $0.03 < d < 2$~km has a power law slope of -1.2, we estimate there should be $4.3 \times 10^{10}$ bodies with $d \ge 0.03$~km.  Accordingly, between 200$\mu$m and 30 m bodies, the SFD has to increase by about 16 orders of magnitude in number over a little more than 5 orders of magnitude in size. The only way to do this is to move away from a shallow power law slope of $q = -1.2$ to one that is much steeper (i.e. $q \sim -3$), with the change taking place at sizes smaller than 30~m.  We find this calculation compelling because our inferred slope for the steep branch is comparable to $q = -2.7$, the slope of the main asteroid belt for $d < 200$~m.

{ Although this suggests that the KBO population is collisionally active and its SFD is at collisional equilibrium, another condition has to be satisfied for the argument to be complete: the two inflection points of the SFD at $d\sim 30$~m and $d\sim 2$~km need to be causally related (O'Brien and Greenberg, 2003). In other words, a KBOs with $d \sim 30$~m needs to be capable of disrupting a KBOs as large as $d \sim 1$-2~km.}

To check whether this is plausible, we assume that a projectile capable of making a crater the same size as the target body will disrupt it. Using the projectile-to-crater scaling law in the gravity regime (\ref{SLg}) with $g=0.05$~mm/s$^{2}$ (i.e. 1/10 of Arrokoth's, { suitable for a km-size target}) we find that a km-sized crater can be formed by a 23~m projectile. 
As added evidence, consider Hayabusa2's Small Carry-on Impactor (SCI) experiment on the 1 km diameter C-type asteroid Ryugu. Here a 2~kg copper plate was shot into Ryugu at 2 km/s, where it made a crater that was 13--17~m in diameter (Arakawa et al. 2020). If we convert the plate mass into a comparable comet projectile (i.e., 0.2~m for { an assumed density of} 500~g~cm$^{-3}$, { suitable for KBO projectiles}), it yields a crater to projectile ratio of $\sim 75$.  Using this ratio, a km-size crater on a km-size target (i.e. a catastrophic collision) could be made by a 13~m projectile. { Thus, both calculations suggest that {10}--30~m projectiles can well disrupt 1~km targets in the Kuiper belt.}  

All of this is an argument for collisional evolution dominating the sub-km KBO population.  In addition, the observation of sub-km craters on Arrokoth is by itself an additional strong argument in favor of collisional equilibrium of the KBO population. In fact, these craters demonstrate that bodies of 10-30~m exist in large number the Kuiper belt. The {\it streaming instability} (Youdin and Goodman, 2005; Johansen et al. 2007) is today the most credited mechanism for the formation of planetesimals, particularly since it has been shown to explain the orientation of KBO binaries (Nesvorny et al., 2019a) and the structure of Arrokoth itself (McKinnon et al., 2020). Bodies of 10-30~m have such a low gravity that is inconceivable they could form by the streaming instability, which is a gravitational instability process (Gerbig et al., 2020; Klahr and Schreiber, 2020). Instead, they are likely to be fragments of larger bodies produced in collisions. If this is the case, then the KBO population should be at collisional equilibrium at these sizes because once bodies of a given size begin to be produced in collisions, collisional equilibrium is rapidly reached (Bottke et al., 2005).

In conclusion, { we claim that the size distribution of KBOs is at collisional equilibrium because: (i) the slope of the size distribution of KBOs in the 30~m--1~km range determined in this work is similar to that of the asteroid belt (which is undoubtedly at collisional equilibrium) in the 200~m--1~km range; (ii) the small strength of KBOs justifies that the shallow portion of their SFD extends down to smaller sizes than for the asteroid population (i.e. 30~m vs. 200~m); (iii) the large abundance of KBO dust detected by the New {Horizons} mission suggests that the Kuiper belt is collisionally active even today; (iv) the very existence of dekameter-sized KBOs, which cannot have been formed as primordial objects by the streaming instability, suggests fragmentation of larger objects.}  { Our results lead to a predicted size distribution of the KBO population (Fig.~\ref{fullSFD}) down to dust sizes, with a characteristic upturn in slope at $d\sim$ {10}--30~m. This prediction will hopefully be testable with future stellar occultation surveys or with in-situ exploration of KBO surfaces at a resolution of $\sim 10$~m.}

\section{Acknowledgments}
We thank the two anonymous reviewers who, through their detailed comments, helped making this paper more clear and complete. 

\section{References}

\begin{itemize}

\item[--] Adams, E.~R., and 6 colleagues 2014.\ De-biased Populations of Kuiper Belt Objects from the Deep Ecliptic Survey.\ The Astronomical Journal 148, 55.

\item[--] Arakawa, M., and 67 colleagues 2020.\ An artificial impact on the asteroid (162173) Ryugu formed a crater in the gravity-dominated regime.\ Science 368, 67.
  
\item[--] Arimatsu, K., Tsumura, K., Usui, F., Shinnaka, Y., Ichikawa, K., Ootsubo, T., Kotani, T., Wada, T., Nagase, K., Watanabe, J.\ 2019.\ A kilometre-sized Kuiper belt object discovered by stellar occultation using amateur telescopes.\ Nature Astronomy, doi:10.1038/s41550-018-0685-8

\item[--] Attree, N., and 54 colleagues 2018.\ Tensile strength of 67P/Churyumov-Gerasimenko nucleus material from overhangs.\ Astronomy and Astrophysics 611, A33.
  
\item[--] Bernstein, G.~M., Trilling, D.~E., Allen, R.~L., Brown, M.~E., Holman, M., Malhotra, R.\ 2004.\ The Size Distribution of Trans-Neptunian Bodies.\ The Astronomical Journal 128, 1364.

\item[--] Beyer, R.~A., and 17 colleagues 2017.\ Charon tectonics.\ Icarus 287, 161.

%\item[--] Bierhaus, E.~B., Chapman, C.~R., Merline, W.~J., Brooks, S.~M., Asphaug, E.\ 2001.\ Pwyll Secondaries and Other Small Craters on Europa.\ Icarus 153, 264.

  \item[--] Bierhaus, E.~B., Dones, L.\ 2015.\ Craters and ejecta on Pluto and Charon: Anticipated results from the New Horizons flyby.\ Icarus 246, 165–182.

%\item[--] Bierhaus, E.~B., and 6 colleagues 2018.\ Secondary craters and ejecta across the solar system: Populations and effects on impact-crater-based chronologies.\ Meteoritics and Planetary Science 53, 638.

\item[--] Bland, M.~T., Singer, K.~N., McKinnon, W.~B., Schenk, P.~M.\ 2017.\ Viscous relaxation of Ganymede's impact craters: Constraints on heat flux.\ Icarus 296, 275.
  
\item[--] Bottke, W.~F., and 6 colleagues 2002.\ Debiased Orbital and Absolute Magnitude Distribution of the Near-Earth Objects.\ Icarus 156, 399.

\item[--] Bottke, W.~F., and 6 colleagues 2005.\ The fossilized size distribution of the main asteroid belt.\ Icarus 175, 111.

\item[--] Bottke, W.~F., Bro{\v{z}}, M., O'Brien, D.~P., Campo Bagatin, A., Morbidelli, A., Marchi, S.\ 2015.\ The Collisional Evolution of the Main Asteroid Belt.\ Asteroids IV 701.

\item[--] Bottke, W.~F. and 14 colleagues 2020.\ Interpreting the Cratering Histories of Bennu, Ryugu, and Other Spacecraft-explored Asteroids.\ The Astronomical Journal 160. doi:10.3847/1538-3881/ab88d3

\item[--] Brownlee, D.~E., Joswiak, D.~J., Love, S.~G., Nier, A.~O., Schlutter, D.~J., Bradley, J.~P.\ 1993.\ Properties of Cometary and Asteroidal IDPs Identified by He Temperature-Release Profiles.\ Meteoritics 28, 332.

\item[--] Canup, R.~M.\ 2005.\ A Giant Impact Origin of Pluto-Charon.\ Science 307, 546.

\item[--] Davidsson, B.~J.~R., and 47 colleagues 2016.\ The primordial nucleus of comet 67P/Churyumov-Gerasimenko.\ Astronomy and Astrophysics 592, A63.

%\item[--] Duncan, M.~J., Levison, H.~F.\ 1997.\ A scattered comet disk and the origin of Jupiter family comets.\ Science 276, 1670.
  
\item[--] Emery, J.~P., Marzari, F., Morbidelli, A., French, L.~M., Grav, T.\ 2015.\ The Complex History of Trojan Asteroids.\ Asteroids IV 203.
  
\item[--] Fuentes, C.~I., George, M.~R., Holman, M.~J.\ 2009.\ A Subaru Pencil-Beam Search for m$_{R}$ \raisebox{-0.5ex}\textasciitilde 27 Trans-Neptunian Bodies.\ The Astrophysical Journal 696, 91.

\item[--] Fraser, W.~C., Brown, M.~E., Morbidelli, A., Parker, A., Batygin, K.\ 2014.\ The Absolute Magnitude Distribution of Kuiper Belt Objects.\ The Astrophysical Journal 782, 100.

\item[--] Gerbig, K., Murray-Clay, R.~A., Klahr, H., Baehr, H.\ 2020.\ Requirements for gravitational collapse in planetesimal formation --- the impact of scales set by Kelvin-Helmholtz and nonlinear streaming instability.\ arXiv e-prints arXiv:2001.10552.
  
\item[--] Gladman, B., Marsden, B.~G., Vanlaerhoven, C.\ 2008.\ Nomenclature in the Outer Solar System.\ The Solar System Beyond Neptune 43.

\item[--] Grav, T., and 16 colleagues 2011.\ WISE/NEOWISE Observations of the Jovian Trojans: Preliminary Results.\ The Astrophysical Journal 742, 40.
  
\item[--] Greenstreet, S., Gladman, B., McKinnon, W.~B.\ 2015.\ Impact and cratering rates onto Pluto.\ Icarus 258, 267.

\item[--] Greenstreet, S., Gladman, B., McKinnon, W.~B., Kavelaars, J.~J., Singer, K.~N.\ 2019.\ Crater Density Predictions for New Horizons Flyby Target 2014 MU69.\ The Astrophysical Journal 872, L5.

\item[--] Grun, E., Zook, H.~A., Fechtig, H., Giese, R.~H.\ 1985.\ Collisional balance of the meteoritic complex.\ Icarus 62, 244.
  
\item[--] Heinze, A.~N., Trollo, J., Metchev, S.\ 2019.\ The Flux Distribution and Sky Density of 25th Magnitude Main Belt Asteroids.\ The Astronomical Journal 158, 232.

\item[--] Holsapple, K.~A., Housen, K.~R.\ 2007.\ A crater and its ejecta: An interpretation of Deep Impact.\ Icarus 187, 345.
  
%\item[--] Housen, K.~R., Holsapple, K.~A.\ 2011.\ Ejecta from impact craters.\ Icarus 211, 856.
  
\item[--] Ivanov, B.~A., Neukum, G., Bottke, W.~F., Hartmann, W.~K.\ 2002.\ The Comparison of Size-Frequency Distributions of Impact Craters and Asteroids and the Planetary Cratering Rate.\ Asteroids III 89.

\item[--] Johansen, A., Oishi, J.~S., Mac Low, M.-M., Klahr, H., Henning, T., Youdin, A.\ 2007.\ Rapid planetesimal formation in turbulent circumstellar disks.\ Nature 448, 1022-1025.

\item[--] Johansen, A., Mac Low, M.-M., Lacerda, P., Bizzarro, M.\ 2015.\ Growth of asteroids, planetary embryos, and Kuiper belt objects by chondrule accretion.\ Science Advances 1, 1500109.

\item[--] Johnson, B.~C., Bowling, T.~J., Trowbridge, A.~J., Freed, A.~M.\ 2016.\ Formation of the Sputnik Planum basin and the thickness of Pluto's subsurface ocean.\ Geophysical Research Letters 43, 10,068.
  
\item[--] Keane, J.~T., and 20 colleagues 2020.\ The Geophysical Environment of (486958) Arrokoth.\ Lunar and Planetary Science Conference 2444.

%\item[--] Kirchoff, M.~R., Schenk, P.\ 2010.\ Impact cratering records of the mid-sized, icy saturnian satellites.\ Icarus 206, 485.

  \item[--] Klahr, H., Schreiber, A.\ 2020.\ Turbulence sets the length scale for planetesimal formation: Local 2D simulations of streaming instability and planetesimal formation.\ arXiv e-prints.

 \item[--] Lawler, S.~M., Kavelaars, J.~J., Alexandersen, M., Bannister, M.~T., Gladman, B., Petit, J.-M., Shankman, C.\ 2018a.\ OSSOS: X. How to use a Survey Simulator: Statistical Testing of Dynamical Models Against the Real Kuiper Belt.\ Frontiers in Astronomy and Space Sciences 5, 14.

\item[--] Marchi, S., Chapman, C.~R., Barnouin, O.~S., Richardson, J.~E., Vincent, J.-B.\ 2015.\ Cratering on Asteroids.\ Asteroids IV 725.
  
\item[--] Michel, P., O'Brien, D.~P., Abe, S., Hirata, N.\ 2009.\ Itokawa's cratering record as observed by Hayabusa: Implications for its age and collisional history.\ Icarus 200, 503.

\item[--] McKinnon, W.~B., and 28 colleagues 2020.\ The solar nebula origin of (486958) Arrokoth, a primordial contact binary in the Kuiper Belt.\ Science 367, aay6620.

%\item[--] Nesvorn{\'y}, D., Janches, D., Vokrouhlick{\'y}, D., Pokorn{\'y}, P., Bottke, W.~F., Jenniskens, P.\ 2011.\ Dynamical Model for the Zodiacal Cloud and Sporadic Meteors.\ The Astrophysical Journal 743, 129.
  
\item[--] Nesvorn{\'y}, D., Vokrouhlick{\'y}, D., Morbidelli, A.\ 2013.\ Capture of Trojans by Jumping Jupiter.\ The Astrophysical Journal 768, 45.
  
\item[--] Nesvorn{\'y}, D.\ 2015a.\ Jumping Neptune Can Explain the Kuiper Belt Kernel.\ The Astronomical Journal 150, 68. 

\item[--] Nesvorn{\'y}, D.\ 2015b.\ Evidence for Slow Migration of Neptune from the Inclination Distribution of Kuiper Belt Objects.\ The Astronomical Journal 150, 73. 

\item[--] Nesvorn{\'y}, D., Vokrouhlick{\'y}, D.\ 2016.\ Neptune's Orbital Migration Was Grainy, Not Smooth.\ The Astrophysical Journal 825, 94. 

\item[--] Nesvorn{\'y}, D., Vokrouhlick{\'y}, D., Dones, L., Levison, H.~F., Kaib, N., Morbidelli, A.\ 2017.\ Origin and Evolution of Short-period Comets.\ The Astrophysical Journal 845, 27. 
  
\bibitem[Nesvorn{\'y} et al.(2018)]{2018NatAs...2..878N} Nesvorn{\'y}, D., Vokrouhlick{\'y}, D., Bottke, W.~F., Levison, H.~F.\ 2018.\ Evidence for very early migration of the Solar System planets from the Patroclus-Menoetius binary Jupiter Trojan.\ Nature Astronomy 2, 878–882.

\item[--] Nesvorn{\'y}, D., Li, R., Youdin, A.~N., Simon, J.~B., Grundy, W.~M.\ 2019a.\ Trans-Neptunian binaries as evidence for planetesimal formation by the streaming instability.\ Nature Astronomy 3, 808.

\item[--]  Nesvorn{\'y}, D., and 11 colleagues 2019b.\ OSSOS. XIX. Testing Early Solar System Dynamical Models Using OSSOS Centaur Detections.\ The Astronomical Journal 158, 132.

 \item[--] Nesvorn{\'y}, D., Li, R., Youdin, A.~N., Simon, J.~B., Grundy, W.~M.\ 2019.\ Trans-Neptunian binaries as evidence for planetesimal formation by the streaming instability.\ Nature Astronomy 3, 808–812.

\item[--] O'Brien, D.~P., Greenberg, R.\ 2003.\ Steady-state size distributions for collisional populations:. analytical solution with size-dependent strength.\ Icarus 164, 334.

\item[--] Petit, J.-M., and 16 colleagues 2011.\ The Canada-France Ecliptic Plane Survey{\textemdash}Full Data Release: The Orbital Structure of the Kuiper Belt.\ The Astronomical Journal 142, 131.

  \item[--] Poppe, A.~R., and 8 colleagues 2019.\ Constraining the Solar System's Debris Disk with In Situ New Horizons Measurements from the Edgeworth-Kuiper Belt.\ The Astrophysical Journal 881, L12.

\item[--] Richardson, J.~E., Melosh, H.~J., Greenberg, R.~J., O'Brien, D.~P.\ 2005.\ The global effects of impact-induced seismic activity on fractured asteroid surface morphology.\ Icarus 179, 325.
  
\item[--] Robbins, S.~J. and 10 colleagues 2014.\ The variability of crater identification among expert and community crater analysts.\ Icarus 234, 109–131. doi:10.1016/j.icarus.2014.02.022 

\item[--] Robbins, S.~J., and 29 colleagues 2017.\ Craters of the Pluto-Charon system.\ Icarus 287, 187.

%\item[--] Schenk, P.~M., Chapman, C.~R., Zahnle, K., Moore, J.~M.\ 2004.\ Ages and interiors: the cratering record of the Galilean satellites.\ Jupiter. The Planet, Satellites and Magnetosphere 427.
  
\item[--] Schlichting, H.~E., Ofek, E.~O., Sari, R., Nelan, E.~P., Gal-Yam, A., Wenz, M., Muirhead, P., Javanfar, N., Livio, M.\ 2012.\ Measuring the Abundance of Sub-kilometer-sized Kuiper Belt Objects Using Stellar Occultations.\ The Astrophysical Journal 761, 150. 

\item[--] Shankman, C., Gladman, B.~J., Kaib, N., Kavelaars, J.~J., Petit, J.~M.\ 2013.\ A Possible Divot in the Size Distribution of the Kuiper Belt's Scattering Objects.\ The Astrophysical Journal 764, L2.

\item[--] Singer, K.~N., and 25 colleagues 2019.\ Impact craters on Pluto and Charon indicate a deficit of small Kuiper belt objects.\ Science 363, 955.

\item[--] Sommerfeld, R.A., 1974. A Weibull prediction for the tensile strength -- volume relationship for snow. J. Geophys. Res. 79, 3353-3356
  
\item[--] Spencer, J.~R., and 77 colleagues 2020.\ The geology and geophysics of Kuiper Belt object (486958) Arrokoth.\ Science 367, aay3999.

  \item[--] Strom, R.~G., Marchi, S., Malhotra, R.\ 2018.\ Ceres and the terrestrial planets impact cratering record.\ Icarus 302, 104.

\item[--] Youdin, A.~N., Goodman, J.\ 2005.\ Streaming Instabilities in Protoplanetary Disks.\ The Astrophysical Journal 620, 459-469. 
  
%\item[--] Wetherill, G.~W.\ 1967.\ Collisions in the Asteroid Belt.\ Journal of Geophysical Research 72, 2429.
\item[--] Weaver, H.~A., and 50 colleagues 2016.\ The small satellites of Pluto as observed by New Horizons.\ Science 351, aae0030.

%  \item[--] Williams, K.~E., Pappalardo, R.~T.\ 2011.\ Variability in the small crater population on Callisto.\ Icarus 215, 253.

\end{itemize}
\end{document}